\documentclass[useAMS,usenatbib]{mn2e}

\usepackage{subfigure}
\usepackage{graphicx}
\usepackage{dcolumn}
\usepackage{bm}
\usepackage{longtable}
\usepackage{amsmath}
\usepackage{mathrsfs}
\usepackage{amsfonts}
\usepackage[usenames]{color}
\usepackage{hyperref}

\usepackage{epstopdf}

\newif\ifAMStwofonts

\newcommand{\be} {\begin{equation}}
\newcommand{\ee} {\end{equation}}

\newcommand{\bea}{\begin{eqnarray}}
\newcommand{\eea}{\end{eqnarray}}
\newcommand{\bdm}{\begin{displaymath}}
\newcommand{\edm}{\end{displaymath}}
\newcommand{\ba} {\begin{array}}
\newcommand{\ea} {\end{array}}

\newcommand{\bfg}  {\begin{figure}}
\newcommand{\efg}  {\end{figure}}
\newcommand{\bfgd}  {\begin{figure*}}
\newcommand{\efgd}  {\end{figure*}}
\newcommand{\incgr} {\includegraphics}
\newcommand{\mbf}{\mathbf}

\newcommand{\btb}{\begin{table}}
\newcommand{\etb}{\end{table}}

\def\fnl {f_{\rm NL}}
\def\ad  {A_{\rm D}}
\def\af  {A_{\rm F}}
\def\as  {A_{\rm S}}

\title[Foregrounds impact on $\fnl$]{Foreground influence on primordial non-Gaussianity estimates: needlet analysis of WMAP 5-year data}
\author[P. Cabella, et al.]
{P. Cabella$^{1}$, 
D. Pietrobon$^{1,2}$, 
M. Veneziani$^{4}$, 
A. Balbi $^{1,3}$, 
R. Crittenden$^{2}$, 
\newauthor G. de Gasperis$^{1}$, 
C. Quercellini$^{1}$  
and N. Vittorio$^{1,3}$\\  
\\
$1$ Dipartimento di Fisica, Universit\`a di Roma ``Tor Vergata'', Rome, Italy\\
$2$ Institute of Cosmology and Gravitation, Dennis Sciama Building Burnaby Road Portsmouth, PO1 3FX United Kingdom\\
$3$ INFN Sezione di Roma ``Tor Vergata''\\
$4$ Dipartimento di Fisica, Universit\`a di Roma ``La Sapienza'', Rome, Italy}

\begin{document}

\label{firstpage}
\maketitle

\begin{abstract}
We constrain the amplitude of primordial non-Gaussianity in the CMB data
taking into account the presence of foreground residuals in the maps.
We generalise the needlet bispectrum estimator marginalizing over the 
amplitudes of thermal dust, free-free and synchrotron templates. We apply 
our procedure to WMAP 5 year data, finding $\fnl= 38\pm 47$  (1 $\sigma$), while the analysis without marginalization provides $\fnl= 35\pm 42$. Splitting the marginalization over each foreground
separately, we found that the estimates of $\fnl$ are positively cross correlated of $17\%, 12\%$ with the 
dust and synchrotron respectively, while a negative cross correlation of about $-10\%$ is found for the 
free-free component.    
\end{abstract}
     
\section{Introduction}
There has been considerable activity recently in constraining
the amount of non-Gaussianity present in CMB data.  This is principally motivated by theoretical interest into deviations of
primordial fluctuations from Gaussian statistics--- a
natural outcome of several implementations of the inflationary
scenario, which can be used as a tool to rule out specific models. More mundanely, non-Gaussianity can also be produced by undetected systematics, which may suggest problems in the dataset. It could also indicate that elements are missing in the standard cosmological model; for example, anomalies which have recently been detected on large scales have been put forward as evidence of cosmic defects or an anisotropic universe.  

The primordial non-Gaussianity in inflationary models is usually
characterised through the introduction of the $\fnl$ parameter (see
e.~g.~
\cite{Luo1994,Heavens1998,SpergelGoldberg1999,KomatsuSpergel2001})
which sets the amplitude of the non-linear contribution with respect
to the leading order:
\begin{equation}
\label{eq:fnl_eq}
\Phi({\bf x})=\Phi_{\rm L}({\bf x})+\fnl(\Phi_{\rm L}({\bf x})^2-\langle\Phi_{\rm L}({\bf x})^2\rangle)
\end{equation}
where $\Phi({\bf x})$ is the primordial gravitational potential and $\Phi_{\rm L}({\bf x})$ is a Gaussian field. 
In the following we will focus on primordial non-Gaussianity of local type described by Eq.~\ref{eq:fnl_eq}.
Different inflationary scenarios predict different values of $f_{\rm
NL}$; while  in the standard slow roll inflationary scenario
\citep{Guth1981,Sato1981,Linde1982,AlbrechtSteinhardt1982},  
$f_{\rm NL}$ it is predicted to be of the order of unity (see also \citep{Maldacena:2003,Acquaviva:2003}), in several alternative models \citep{LythWands2002,LindeMukhanov2006,AlabidiLyth2006,Mizuno2008,Khoury2002PhDT,SteinhardtTurok2002,LehnersSteinhardt2008} it can take much larger values (see e.g. \cite{Bartolo2004NGreview} for a review). 
Since the release of the first year WMAP data \citep{Bennett:2003ca, Komatsu:2003}, there has been a drastic reduction of the upper  limits of $\fnl$. The most recent constraints coming from WMAP data  using different techniques can be found in \cite{Komatsu2008wmap5}, \cite{Smith2009fnlFore}, \cite{YadavWandelt2007}, \cite{Curto2009NG}, \cite{Curto2008waveNG}, \cite{Pietrobon2008NG}, \cite{Rudjord2009needBis}, \cite{Rudjord:2009au}, \cite{Smidt:2009ir} and \cite{VielvaSanz:2009}.
Results on $f_{\rm NL}$ with suborbital experiments can be found in \citep{B2kNG,Curto2008Archeops}.   
Interestingly, some of these \citep{YadavWandelt2007,Rudjord:2009au}
reported a non-null detection of  $\fnl$ at more than 2$\sigma$
level in the WMAP data, leading to a debate on the significance of the signal and whether it may have a cosmological origin
or be due to residual foreground or instrumental contamination \citep{Smith2009fnlFore}. 

Non-Gaussianity is also
expected from unremoved contamination from
astrophysical sources, or foregrounds. When component separation
techniques are applied to CMB data (e.g.\
\cite{Maino:2002FastICA,Tegmark:2003,Bonaldi:2007,Leach:2008, Bernui:2009}),
residual foregrounds remain, and while they are subdominant, they can still be a source of non-Gaussianity, and this could be confused with a primordial signature and affect the constraints on $f_{\rm NL}$.  

In this paper, which is complementary to our previous work \citep{Pietrobon2008NG}, we aim at generalising the needlet bispectrum estimator to the case in which such foreground residuals are present in the data. Another marginalization technique, using the fast cubic estimator on the bispectrum data, can be found in \cite{Smith2009fnlFore}. 

The paper is organised as follows. In section \ref{sect:formalism} we give a brief introduction of the needlets and their bispectrum; section \ref{sect:estimator_gen} addresses the generalisation of the needlet bispectrum estimator in presence of foreground residuals, in section \ref{sect:dataset} we describe the CMB and the foregrounds dataset used and show our results, in Section \ref{sect:concl} we comment on our findings.  
\section{Needlet formalism and general applications}
\label{sect:formalism}
Here we give a brief summary of the needlet formalism, but one should refer to \cite{Marinucci2008} and references therein for a
more complete discussion. The mathematical approach is described in \cite{NarcowichPetrushevWard2006}, \cite{Baldi2006}, \cite{Baldi2007} and \cite{Baldi2008Adaptivedensity}.

Needlets are filter functions with some appealing properties; in particular they are well localised both in real and harmonic space, and are straight forward to implement in practice.   The analytical function describing a needlet is:
\be
  \label{eq:need_def}
  \psi _{jk}(\hat\gamma) = \sqrt{\lambda_{jk}}
  \sum_{\ell} b\Big(\frac{\ell}{B^{j}}\Big)
  \sum_{m=-\ell}^{\ell}\overline{Y}_{\ell m}(\hat\gamma)Y_{\ell m}(\xi _{jk}).
\ee
Here, $\hat\gamma$ is the direction of a given point on the sphere, $\{\xi_{jk}\}$ are the cubature points, weighted 
by $\lambda_{jk}$, for a given spatial frequency $j$ and location $k$ and $b(\cdot)$ is a filter function defined
in harmonic space. $B$ is a free parameter which, once fixed,
determines the width of  $b(\cdot)$ and determines the range of
multipoles which appears in Eq.~\ref{eq:need_def}. Its value is
set according to the angular scales one is interested in exploring. It
can be shown (see {\cite{Marinucci2008}) that 
assuming certain properties of the function $b(\cdot)$ (e.g. finite support in $\ell$-space and
smoothness), the needlets are quasi-exponentially localised around
the $\hat\gamma$ direction \citep{NarcowichPetrushevWard2006}. 
These properties allow needlets to be used as a scale dependent analyser
in both the harmonic and pixel domains. 
 For the reader interested in the details of a practical implementation of needlets using 
Healpix package\footnote{http://healpix.jpl.nasa.gov}
 \citep{Gorski2005Healpix}, we refer to \cite{Pietrobon2006ISW}. 
From the operational point of view, the quantities of interest are the
coefficients of the expansion on the needlets basis defined as:
\bea
  \label{eq:need_coef}
  \beta _{jk} &=&\int_{S^{2}}T(\hat\gamma)\psi _{jk}(\hat\gamma)d\Omega  \nonumber \\
   &=&\sqrt{\lambda_{jk}}\sum_{\ell}b\Big(\frac{\ell}{B^{j}}\Big)\sum_{m=-\ell}^{\ell}a_{\ell m}Y_{\ell m}(\xi _{jk}).
\eea
These coefficients can be used for reconstructing a map as:
\begin{equation}
T(\hat\gamma)\equiv \sum_{j,k}\beta _{jk}\psi _{jk}(\hat\gamma)
\label{recfor}
\end{equation}
As with other varieties of wavelets, needlets allow a multi-scale analysis, which is  
useful for many kinds of data analyses, such as denoising (\cite{Sanz1999Denoising}), point source extraction (see for instance, \cite{Cayon2000PointS,Vielva2003}), asymmetry studies \citep{Cruz2007ColdSpotW3,Pietrobon2008AISO,McEwen2008,Vielva2007ColdSpot,Mcewen:2008} 
and for cross correlating different datasets, as for example CMB and
large scale surveys for the detection of the Integrated Sachs-Wolfe
effect, which sets constraints on dark energy  \citep{Pietrobon2006ISW,Vielva:2006}.
From the coefficients of Eq.~\ref{eq:need_coef} we can estimate the binned CMB power spectrum:
\be
\label{eq:need_ps}
\beta_j \equiv \sum_{jk}\beta_{jk}^2= \sum_\ell b_{\ell,j}^2 \frac{2\ell+1}{4\pi}\mathcal{C}_\ell
\ee
(see \cite{Fay2008PS} and \cite{Pietrobon2006ISW,Pietrobon2008AISO} for
an application to real data).
It is straightforward to generalise this formalism to higher order statistics like
the needlet bispectrum \citep{Lan2008NeedBis,Pietrobon2008NG,Rudjord2009needBis,Pietrobon:2009qg,Rudjord:2009au}:
\be
\label{eq:needbisp}
S_{j_1j_2j_3} = \frac{1}{\tilde N_{\rm p}}\sum_{\rm k}\frac{\beta_{j_1 k}\beta_{j_2 k}\beta_{j_3 k}}{\sigma_{j_1}\sigma_{j_2}\sigma_{j_3}}.
\ee
Here $k$ spans over the pixels of the region of interest, $N_p$ is the total number of these pixels and $\sigma_{j}$ is the standard 
deviation of the needlet coefficients. Eq.~\ref{eq:needbisp} has
been applied to constrain $f_{\rm NL}$ with a needlet analysis, since
it has been proven
to be proportional to the reduced bispectrum
\citep{KomatsuSpergel2001}, and thus to the non-linear
parameter $\fnl$ \citep{Pietrobon2008NG,Rudjord2009needBis,Pietrobon:2009qg,Rudjord:2009au}:
\be
S_{j_1j_2j_3} \propto \fnl \sum_k (\beta^{\rm NG}_{j1 k}\beta^{G}_{j2 k}\beta^{G}_{j3 k} + \rm{perms}). 
\label{eq:fnl_bisp}
\ee
The needlets bispectrum has several useful properties. It can be shown that, given the localisation properties of needlets,
their bispectrum is not heavily affected by the correlation introduced by the masking procedure. Another advantage is that
there is no need to calculate Wigner coefficients (as a pure
bispectrum analysis requires), which is a time consuming step, especially
when high multipoles are considered. These properties translate into a good detection power in constraining primordial non-Gaussianity
(see \citep{Pietrobon2008NG,Rudjord2009needBis,Rudjord:2009au})
comparable with the fast cubic estimator introduced by \cite{KomatsuSpergelWandelt2005} and applied in \cite{Komatsu2008wmap5} and \cite{YadavWandelt2007}.


\section{Method}
\label{sect:estimator_gen}
In this section we generalise the $\fnl$ estimator from the needlet
bispectrum \citep{Pietrobon2008NG} in presence of residual foregrounds.

When constraining primordial non-Gaussianity, one can estimate $\fnl$ by minimising the chi-square:
\be
\chi^2(\fnl)=Y^T\mbf{C}^{-1}Y,
\label{eq:chi2}
\ee
where $Y = Y^{obs}-\langle Y(\fnl)\rangle$ is the difference between an ordered array of data ($Y^{obs}$) and the corresponding theoretical prediction $Y(\fnl)$.  The data can take many forms, such as, the values of Minkowski functionals for different thresholds \citep{Hikage2006MinkFunc}, the densities of valleys or hills when using the local curvature \citep{Cabella2005LocCurv}
or, as in this case, the values of the needlet bispectrum \citep{Pietrobon2008NG,Rudjord2009needBis,Rudjord:2009au}.
In the presence of the expected weak non-Gaussianity, the covariance matrix $C$ can be calculated via Gaussian simulations with the same power spectrum as the observed data.

It has been shown \citep{Pietrobon2008NG,Rudjord2009needBis} that an
unbiased estimator for $\fnl$ is given by
\be
\fnl = \frac{\sum_{\mu\mu^\prime}S^{\rm obs}_\mu\mbf{C}^{-1}_{\mu\mu^\prime}S^{\rm th}_{\mu^\prime}}{\sum_{\mu\mu^\prime}S^{\rm th}_\mu\mbf{C}^{-1}_{\mu\mu^\prime}S^{\rm th}_{\mu^\prime}},
   \label{eq:fnl_estimator}
\ee
where $\mu$ runs over the triplets $\{j_1j_2j_3\}$ and $S^{\rm th}_{\mu}$ represents the ensemble average 
of primordial non-Gaussian realisations ($\fnl =1)$.
Although the process of foreground reduction could make things more complicated, we can make the minimal assumption that 
the final map, contaminated by foreground residuals, can be modeled as:
\bea
T^{\rm sim}(\hat\gamma) &=& T^{\rm G}(\hat\gamma) + \fnl T^{\rm NG}(\hat\gamma) + N(\hat\gamma) \nonumber \\
&+& \alpha_{\rm D}\,D_{\rm}(\hat\gamma)+ \alpha_{\rm F}\,F_{\rm}(\hat\gamma)+ \alpha_{\rm S}\,S_{\rm}(\hat\gamma),
\label{eq:th_map}
\eea
where $D$, $F$ and $S$ are the thermal dust, free-free emission and synchrotron radiation maps, respectively 
and the noise map $N$ is assumed to be Gaussian.
Calculating the needlets coefficients from  Eq.~\ref{eq:th_map} we obtain 
\be
\beta_{jk} = \beta_{jk}^{\rm G} + \beta_{jk}^{\rm N} + \fnl\beta_{jk}^{\rm NG} +
\alpha_{\rm D} \beta_{jk}^{\rm D} +  \alpha_{\rm F} \beta_{jk}^{\rm F} + \alpha_{\rm S} \beta_{jk}^{\rm S}.
\label{eq:th_beta}
\ee
In Figure \ref{fig:templates} we show the needlet coefficients in the
case of $B=2$ and $j=6$ for the three foregrounds templates -- dust, free-free and
synchrotron -- once they have been masked. The maps were converted in
thermodynamic temperature for each channel and combined to form one
single map. Each template was divided by a factor 10, which is the
level of residuals expected.
This factor was estimated in harmonic space through a Monte Carlo Markov Chain algorithm. 
The foreground reduced map used in this analysis has been produced by optimally 
combining the W, V and Q WMAP bands as described below (Eq.~\ref{eq:map_constr}.) 
\begin{figure}
\center
\incgr[width=.4\columnwidth, angle=90]{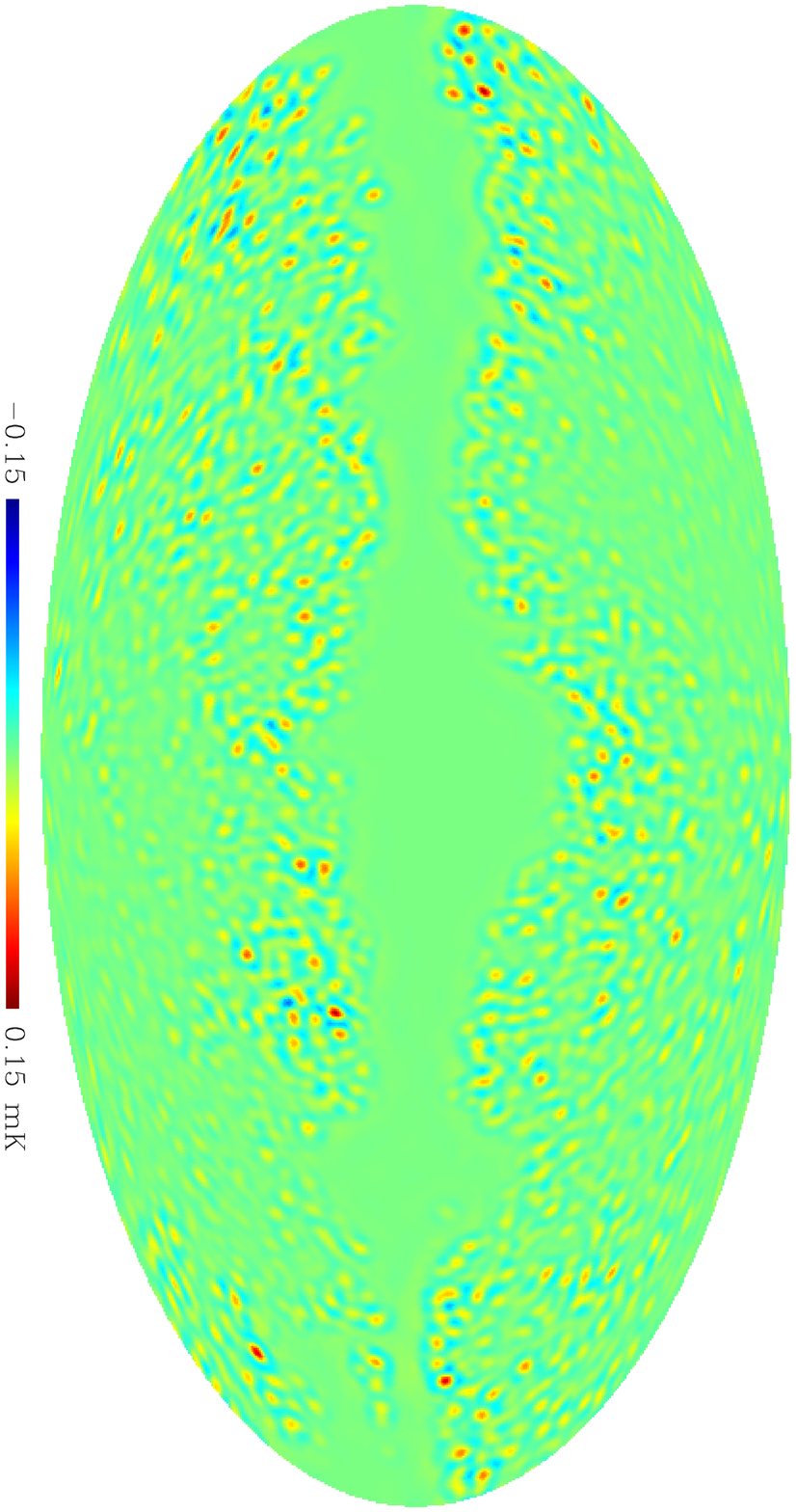}
\incgr[width=.4\columnwidth, angle=90]{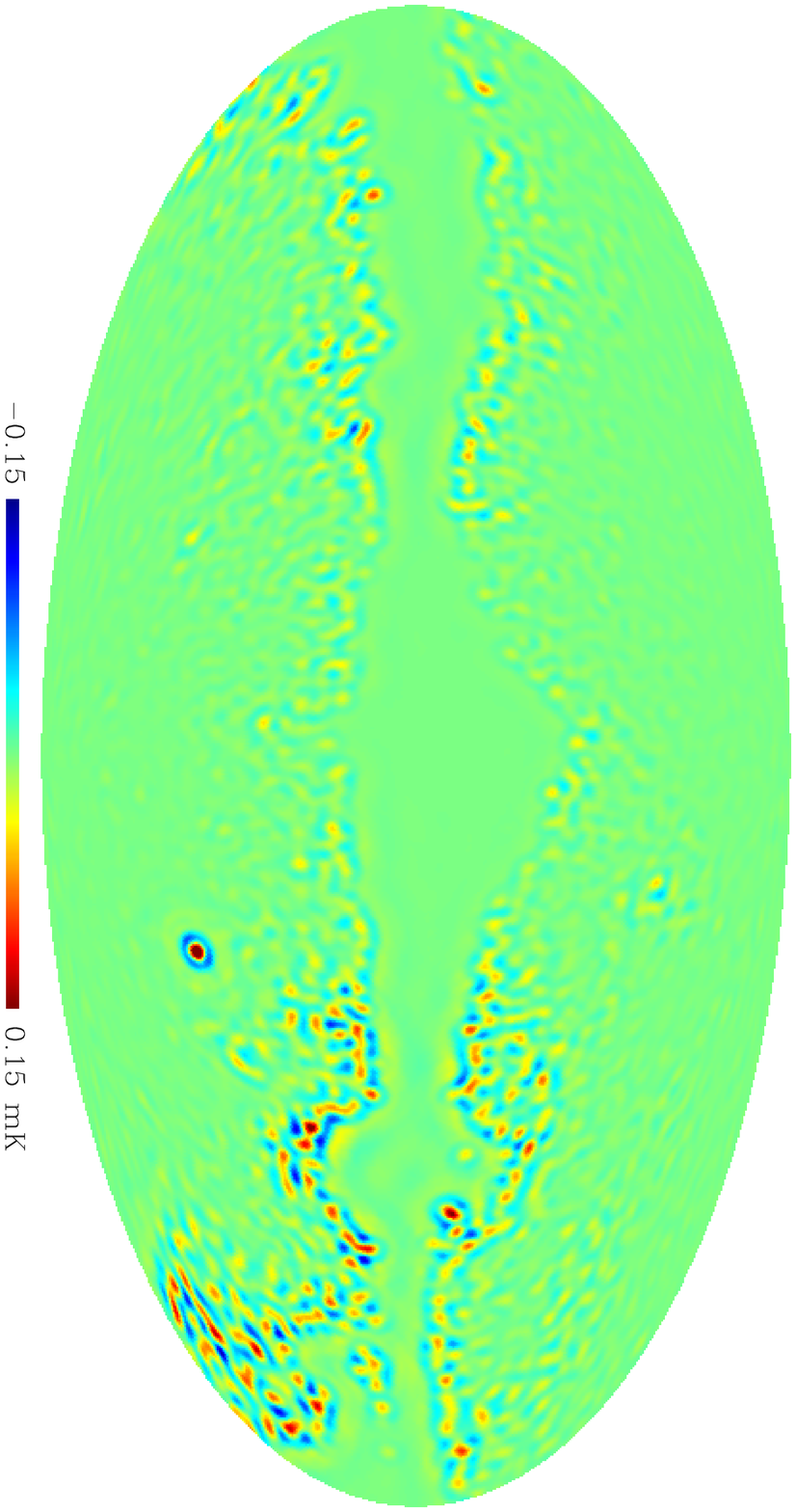}
\incgr[width=.4\columnwidth, angle=90]{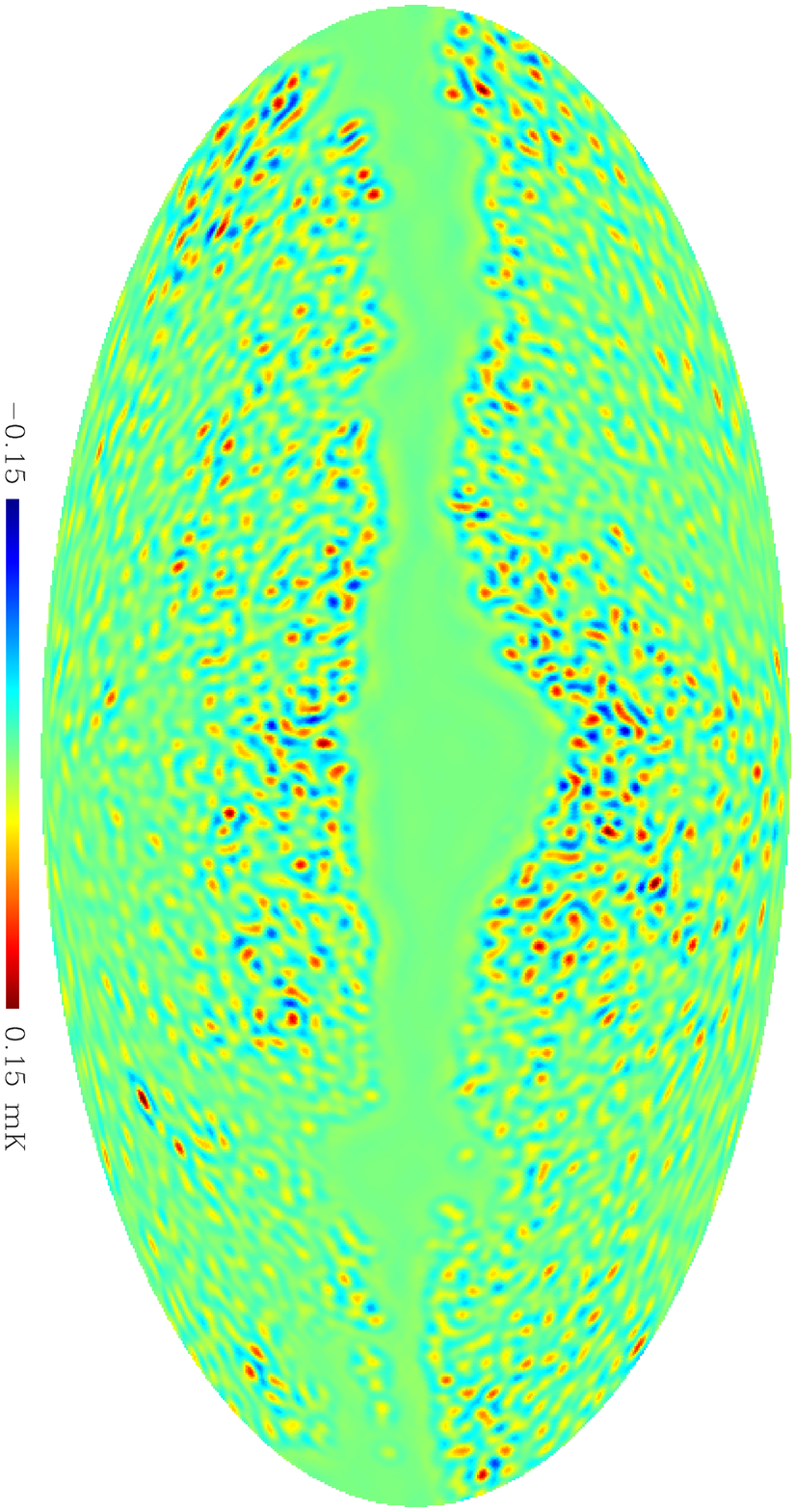}
\incgr[width=.4\columnwidth, angle=90]{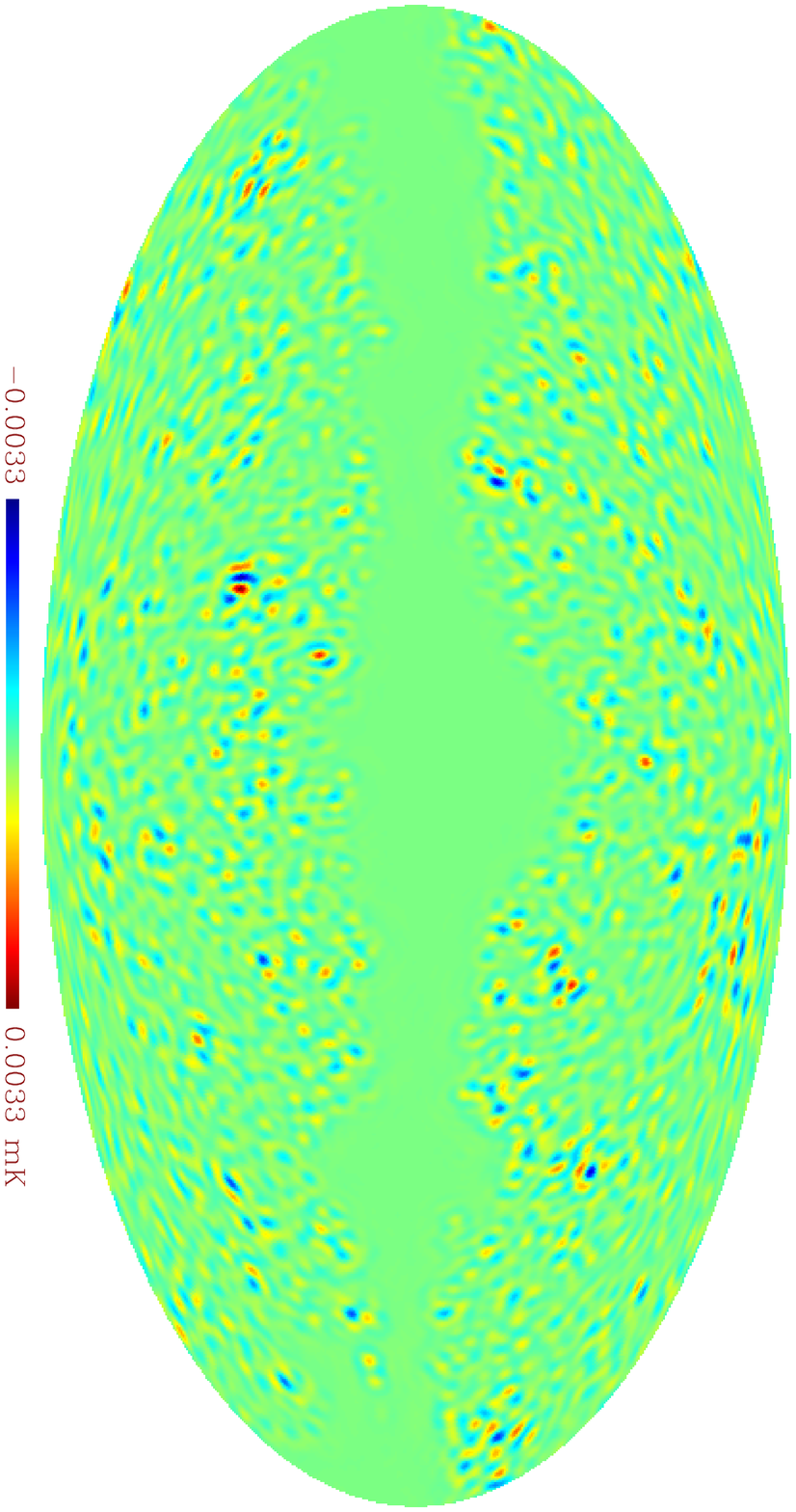}
\caption[CMB and foreground residuals maps]{\small The needlets coefficients with $B=2$ and $j=6$, which are 
sensitive to $64 < \ell < 127$, or angular scales of a few degrees.  The maps shown are for thermal
dust (top), free-free, synchrotron  and the assumed primordial 
non-Gaussian signal for $\fnl=50$ (bottom). The  Kq$75$ mask has been applied.}
\label{fig:templates}
\end{figure}

The angular power spectrum of this optimal CMB map
has been compared to a linear combination of synchrotron, free-free, Galactic dust, CMB and noise.
Synchrotron, free-free and Galactic dust maps are estimated from the data using exactly the same combination in Eq.~\ref{eq:map_constr},
while CMB and noise are estimated through simulations for each band and then combined in the same way as data.
100 realisations of CMB maps have then been generated in order to take into
account the uncertainty due to the cosmic variance. 100 realisations of noise have been created
with variance consistent with the data and then fitted
the amplitude coefficients $p_\gamma$ of each component
$\gamma$, such that 
\be\label{es:lin_comb}
\mathcal{C}_\ell^{tot} = \mathcal{C}_\ell^{\rm CMB}+\sum_\gamma p_\gamma
\mathcal{C}_\ell^\gamma+N_\ell,
\ee
where $N_\ell$ is the detector noise (a similar procedure has been applied in \cite{MarcellaSZ}). 
We find $p_{\rm synch} < 0.1 \times 10^{-1}$,
$p_{\rm dust} < 1.0 \times 10^{-1}$ and $p_{\rm ff} < 0.6 \times
10^{-1}$. The coefficients $p_{\rm \gamma}$ provide an indication of
the power spectrum contamination percentage by the foreground
residuals. The effect on the bispectrum is estimated to be of
the order of $\sim10^{-3}$. 

We can compute the needlet bispectrum expected from the model of the model of residuals described in 
 Eq.~\ref{eq:th_map}, by substituting Eq.~\ref{eq:th_beta} into the definition of the bispectrum.
In principle, this will include many terms including triple products of the various contributions.  
On average, the product of the Gaussian contributions should be zero, and the lowest order contribution from 
the primordial non-Gaussianity is that shown in Eq.~\ref{eq:fnl_bisp}.  We also expect on average that cross terms 
like $\beta^{\rm NG}\beta^{\rm I_1}\beta^{\rm I_2}$ (I=$\{$D,F,S$\}$ will also be zero, as the foregrounds should be 
uncorrelated with the primordial signal.  

However, cross terms which include different foregrounds, of the form $\beta^{\rm I_1}\beta^{\rm I_2}\beta^{\rm I_3}$, 
could potentially be non-zero, particularly if there are physical reasons to expect correlations. (For example, both 
synchrotron and free-free are sensitive to the free electron density.)  Even if such terms are zero on average, they will 
not be zero for specific realisations of the foregrounds; indeed if we think we know the foreground templates, we can 
calculate these terms directly.   For simplicity here, we ignore such terms and focus only on the 'auto-bispectra.'   This has the advantage 
that solving for the best fit amplitudes involves solving a simple set of linear equations.  (Cross terms make the equations non-linear, 
requiring a numerical solution.)  

With this assumption, the theoretical needlet bispectrum ($S$) in presence of foreground residuals can be written as:
\be
S_{j_1j_2j_3} = \fnl S^{\rm NG}_{j_1j_2j_3} + A_{D} S^{\rm
D}_{j_1j_2j_3} +A_{F}S^{\rm F}_{j_1j_2j_3} + A_{S} S^{\rm S}_{j_1j_2j_3},
\label{eq:gen_bis}
\ee
where
\bea
&& S^{\rm NG}_{j_1j_2j_3} = \sum_k \frac{\beta_{j_1k}^{\rm
G}\beta_{j_2k}^{\rm G}\beta_{j_3k}^{\rm
NG}}{\sigma_{j_1}\sigma_{j_2}\sigma_{j_3}} + {\rm perms} \nonumber ;\\
&& S^{\rm I}_{j_1j_2j_3} = \sum_k \frac{\beta_{j_1k}^{\rm
I}\beta_{j_2k}^{\rm I}\beta_{j_3k}^{\rm
I}}{\sigma_{j_1}\sigma_{j_2}\sigma_{j_3}} .
\eea
For ease of notation, we have defined the foreground amplitudes as $A_{\rm I} = \alpha^3_{\rm I}.$  

In Figure \ref{fig:ampl1} we show the bispectrum for the primordial non-Gaussianity with $\fnl=50$ and for the foreground 
templates of WMAP data normalised according to the prescription discussed previously. The bispectrum is shown as a function of the quantity
$X=1/j_1j_2+1/j_1j_3+1/j_3j_2$, which orders them roughly according to their variance (as can be seen in Figure \ref{fig:databisp}.) 

\bfg
\center
\incgr[width=\columnwidth]{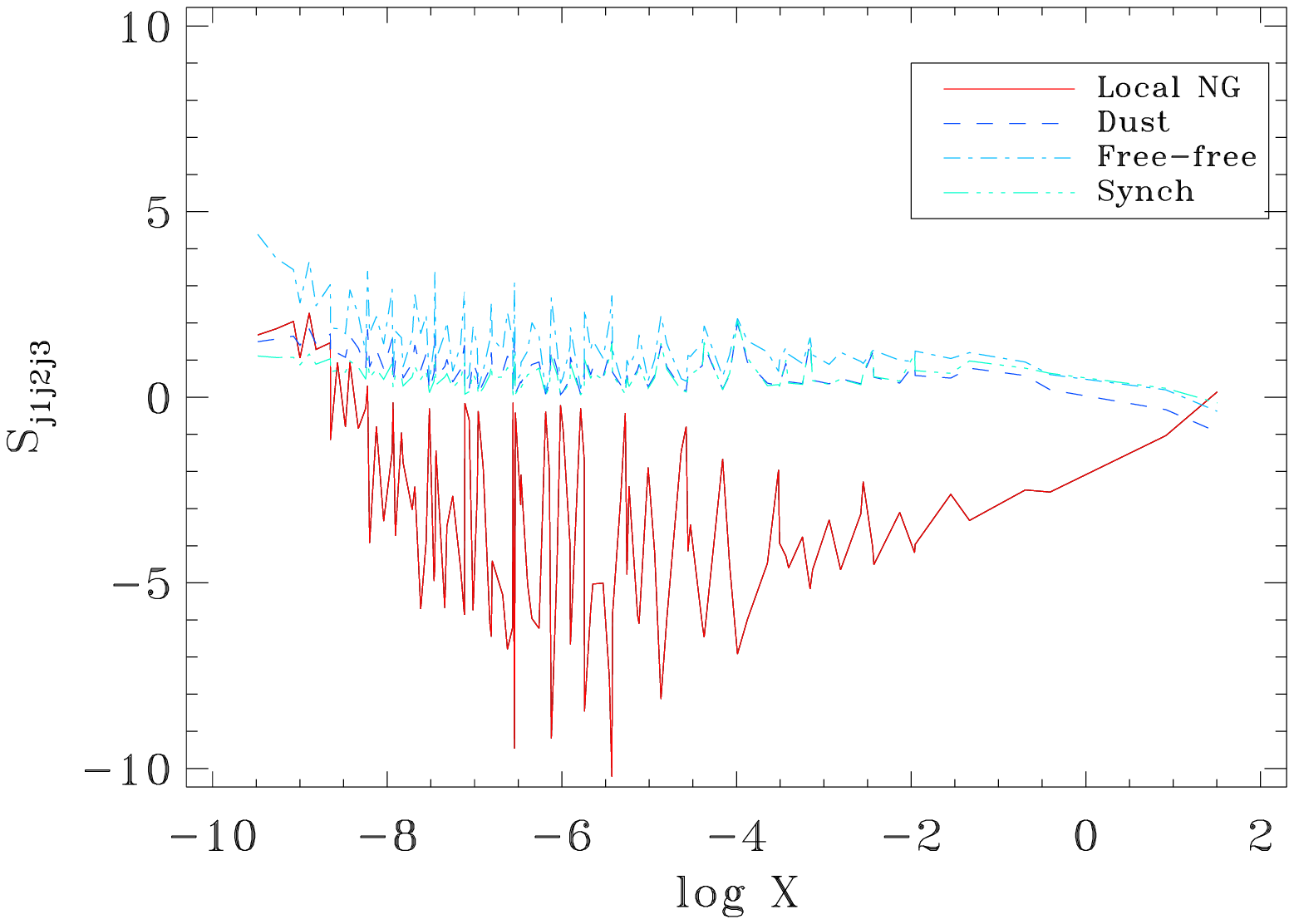}
\caption[Bispectrum templates]{\small Bispectrum amplitudes for the
primordial non-Gaussian signal ($\fnl=50$), and the three WMAP data foreground templates suppressed by a factor 1000, roughly corresponding to 
the expected signal of the residual foregrounds.}
\label{fig:ampl1}
\efg
The $\chi^2$ expression (Eq.~\ref{eq:chi2}) can be easily generalised as follows:
\bea
\chi^2&=&\big(S^{obs}_{\mu}-S_\mu(\fnl)-S^i_\mu(A_i)\big)C_{\mu\mu'}^{-1}\nonumber \\
&&\big(S^{obs}_{\mu'}-S_{\mu'}(\fnl)-S^i_{\mu'}(A_i)\big)^T,
\label{eq:chi2gen}
\eea
where $i$ refers to the i-th foreground template with the
Einstein summation convention assumed, and $S^i_\mu$ is the needlet bispectrum of the given foreground.
Minimising  Eq.~\ref{eq:chi2gen} with respect to $x=(\fnl,\ad,\af,\as)$ we obtain: 
\bea
&& \fnl = \frac{S_i^{\rm NG}C_{ij}^{-1}S_j^{obs}}{S_i^{\rm
NG}C_{ij}^{-1}S_j^{\rm NG}} - \sum_K \frac{S_i^{\rm NG}C_{ij}^{-1}S_j^{\rm K}}{S_i^{\rm
NG}C_{ij}^{-1}S_j^{\rm NG}} A_{\rm K} \nonumber \\
&& A_{\rm I} = \frac{S_i^{\rm I}C_{ij}^{-1}S_j^{obs}}{S_i^{\rm
I}C_{ij}^{-1}S_j^{\rm I}} - \sum_K \frac{S_i^{\rm I}C_{ij}^{-1}S_j^{\rm K}}{S_i^{\rm
I}C_{ij}^{-1}S_j^{\rm I}} A_{\rm K} 
\label{eq:est_solutions}
\eea
where in each case the sum implicitly goes over the other signals $ K=\{ \rm NG,  D, F,\rm S \}$. 

The solution of the previous system provides us with the estimates of $f_{\rm NL}$ with the needlet bispectrum 
in presence of foreground residuals. In the following we present data and simulations 
where this estimator has been applied.

\section{Data set, simulations and results}
\label{sect:dataset}
In the following, the needlets of the simulations and data will be calculated for B=2 unless specified otherwise.  
We used the publicly available WMAP 5-year
data\footnote{http://lambda.gsfc.nasa.gov/}
\citep{Hinshaw2008WMAP5}. The CMB map has been obtained 
combining the foreground reduced maps for the channel Q,V,W according
to:
\be
\label{eq:map_constr}
T^{\rm obs}(\hat{\gamma}) = \sum_{\rm ch} \frac{{n_{\rm obs}(\hat\gamma)}}{\sigma^2_{ch}}T^{\rm red}_{\rm ch}(\hat\gamma),
\ee
where $n_{\rm obs}(\hat\gamma)$ is the number of hits for a given point on the sphere $(\hat\gamma)$,  $\sigma_{ch}$
is the nominal sensitivity for the relative channel $ch$ and the superscript $red$ denotes the reduced dataset.  
We considered  the  WMAP data templates of dust,
free-free and synchrotron emission at the resolution of $N_{\rm side}=256$ corrected by 
the conversion factors from antenna to thermodynamic temperature of
the respective channel; we then combined them as in
Eq.~\ref{eq:map_constr} to have one single map for each template. The data
and foreground templates were then masked with
Kq$75$ mask (downgraded to the same resolution as well) which covers
the dominant Galactic emission over roughly the $30\%$ of the sky.
We obtained our final constraints on $f_{\rm NL}$ by applying the
estimator in its improved fashion to the WMAP 5-year data where:

\begin{itemize}
\item the covariance matrix $\mbf{C}^{-1}$ was calibrated over 20,000 Gaussian simulations;   
\item the needlet foreground bispectra $S_{\rm D},S_{\rm F},S_{\rm S}$ were calculated on the templates described above;  
\item the primordial needlet bispectrum $S^{\rm NG}$ was calculated using
over 100 primordial non-Gaussian maps \citep{Liguori2007NGMaps} convolved with the beams and combined as done for the Gaussian simulations (see also \cite{ElsnerWandelt:2009} for an alternative algorithm to generate temperature and polarization primordial non Gaussian maps).  
\end{itemize}

Figure \ref{fig:databisp} shows the data we used together with the
average and the standard deviation derived from simulations. The error bars on $\fnl$ were computed through the distribution of its estimates for 20,000 Gaussian realizations.
\bfg
\center
\incgr[width=\columnwidth]{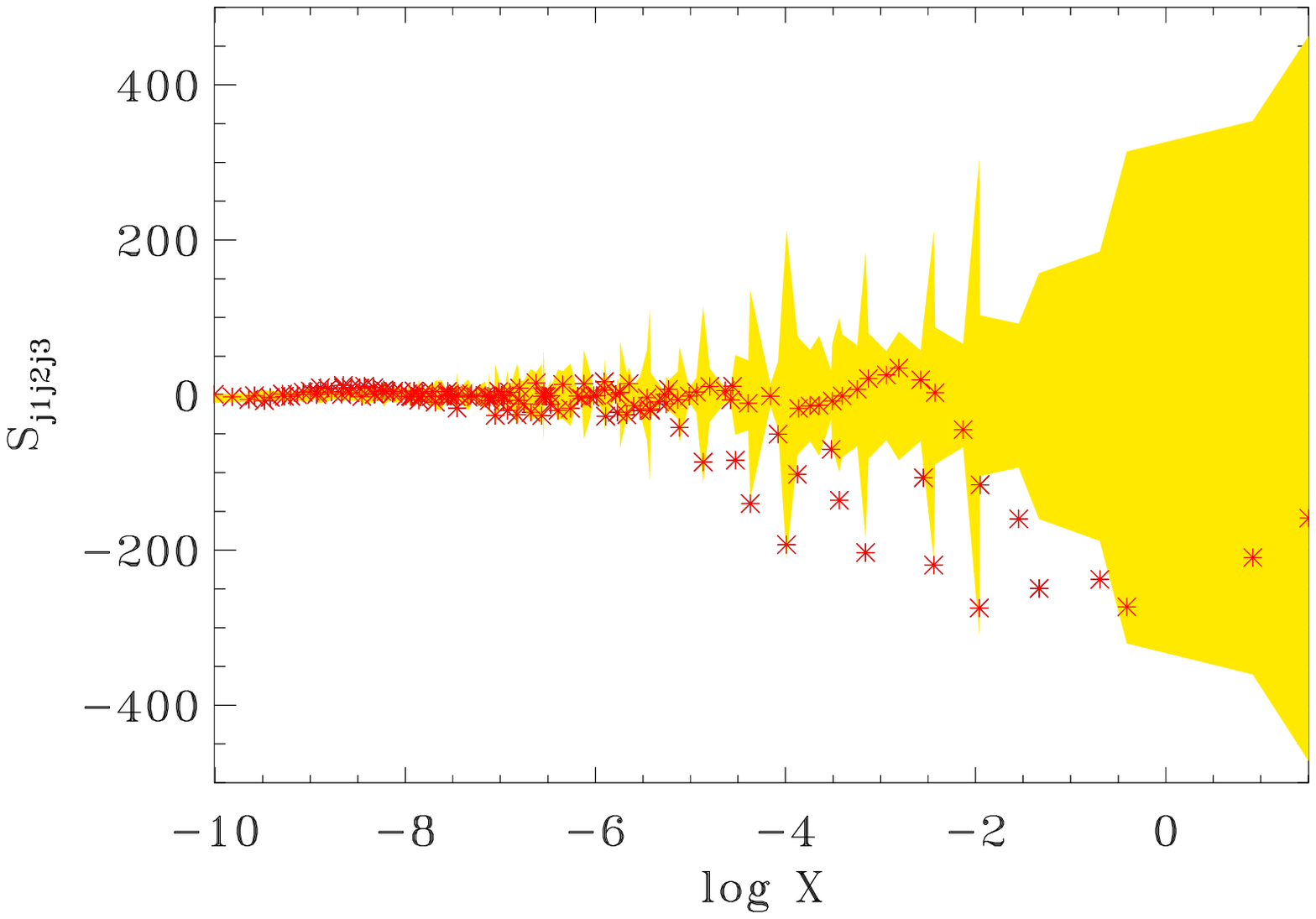}
\caption[Needlet bispectrum measured from WMAP5]{ Needlets bispectrum measured from the WMAP 5-year temperature map. 
The shaded region marks the 68\% confidence level computed from 20,000 Gaussian realisations.}
\label{fig:databisp}
\efg
In Figure \ref{fig:like_fnl_marg} we show our results, where in each
panel we show the estimate of $\fnl$
with and without marginalizing over the foreground templates. 
Without marginalizing, we found  $\fnl=38\pm 42,85$ at 1$\sigma$ and 2$\sigma$, respectively. 
The marginalization over  all the foregrounds brings the constraint to $\fnl=41\pm 47,95$ 
with an increase of the error bars of about $10\%$ and
a positive shift of the mean value. Although the larger error bars mean that we should not attach too much significance to the higher $\fnl$ value, it would be interesting to investigate whether this could partially explain some positive detection present in literature (e.~g.~
\cite{YadavWandelt2007,Rudjord2009needBis}).
Further information can be obtained by looking
at the estimates of $\fnl$ marginalizing over the three foregrounds separately. The enlargement of the error bars due to
each foreground is of the same order of magnitude, but the shift of
the estimate seems to be mostly due to the dust component.
\begin{figure*}[hbtd]
\center
\incgr[width=.85\columnwidth]{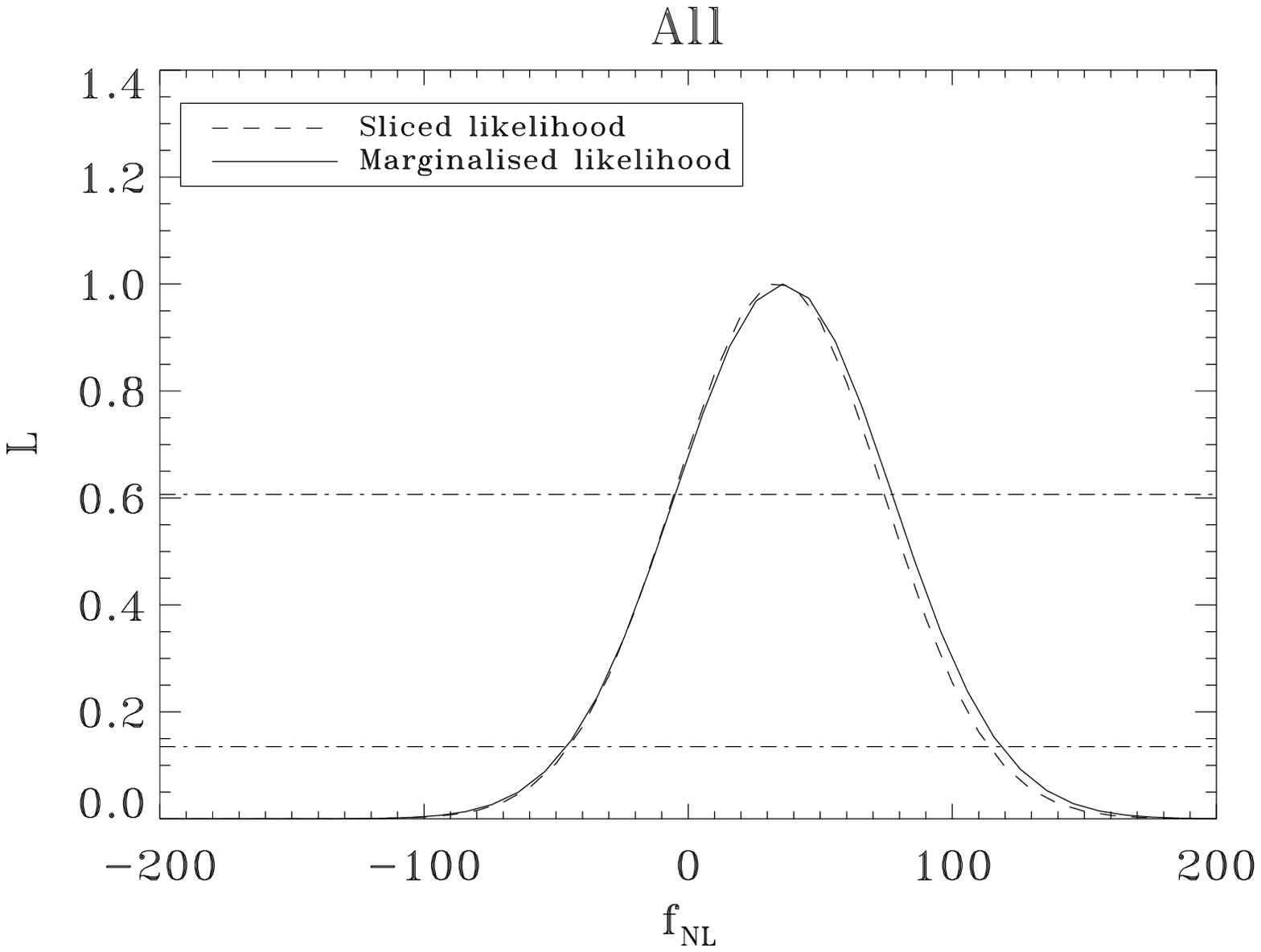}
\incgr[width=.85\columnwidth]{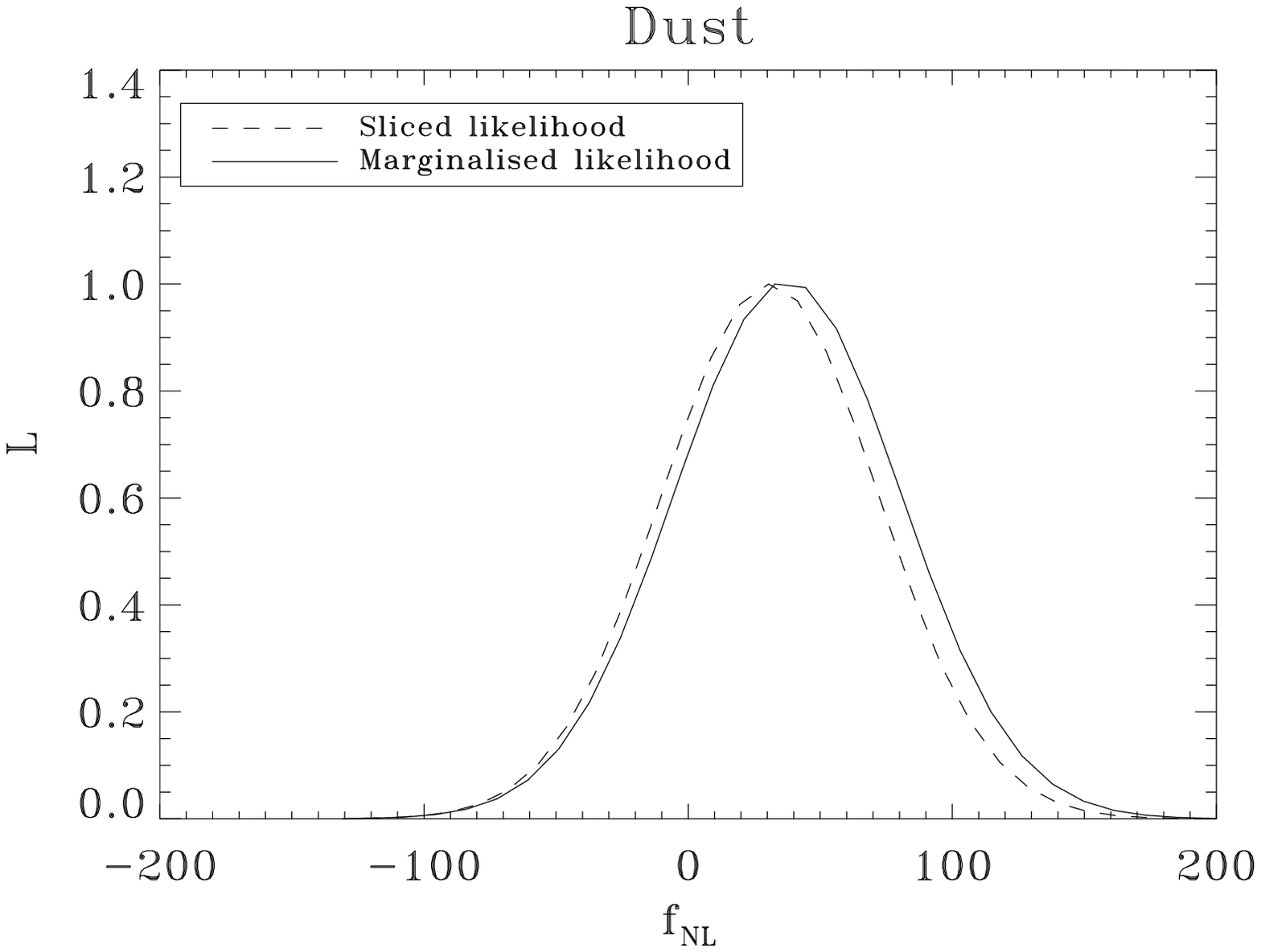}
\incgr[width=.85\columnwidth]{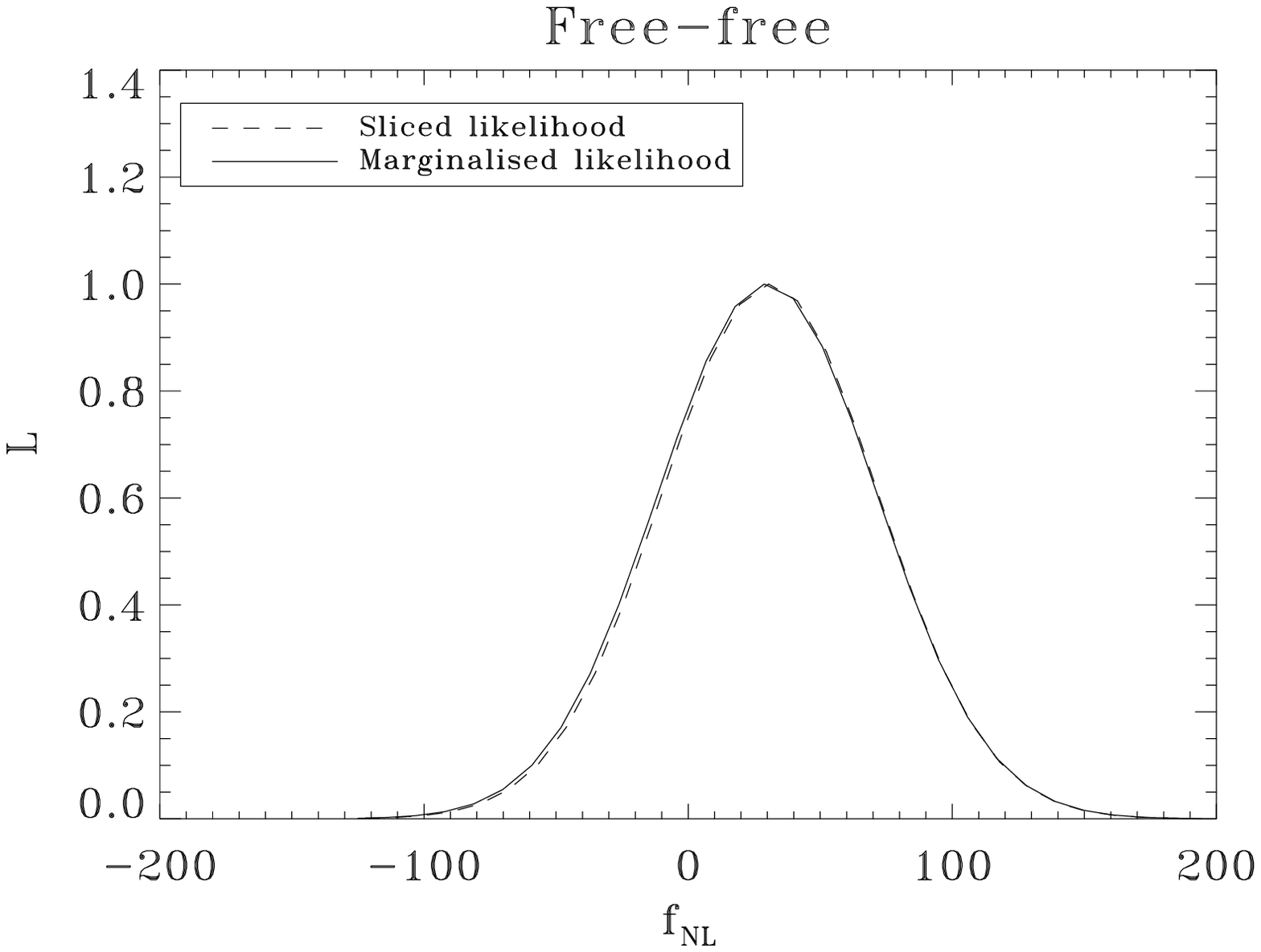}
\incgr[width=.85\columnwidth]{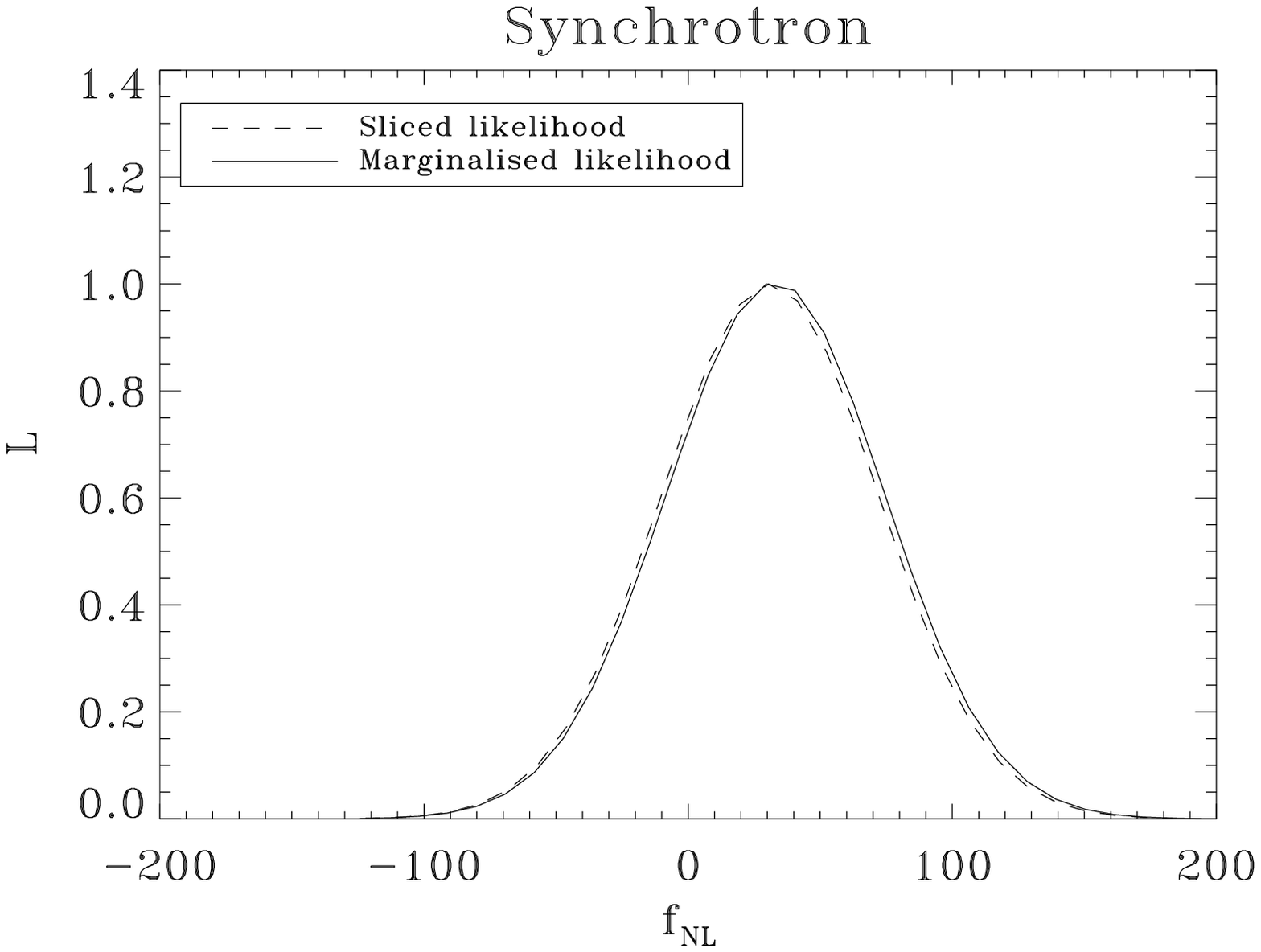}
\caption[Unidimensional likelihood for the $\fnl$
parameter]{ One dimensional likelihoods for the $\fnl$
parameter. Solid lines represent marginalized likelihoods, whereas the
dashed ones refer to the slice $A_{\rm foreg}=0$. The top left panel
shows the analysis including all the three foregrounds, while the
others are derived when a single foreground --- dust (top right), free-free (bottom left) and
synchrotron (bottom right) emission -- is allowed to vary.}
\label{fig:like_fnl_marg}
\end{figure*}

As a further check we carried out a Fisher analysis.  (For a quick-start guide to Fisher matrix see \cite{Coe:2009FisherM}.)
 So far we applied a Monte Carlo approach to find a good estimate
of $\fnl$ in presence of foreground contamination and quantify the
scatter around this evaluation. 

If the likelihood is close to Gaussian, the scatter can be well approximated by the Fisher approach, as 
the Cramer-Rao bound $\sigma_{\theta}\ge 1/(F^{-1})^{1/2}_{ii}$ becomes close to an equality.
The Fisher matrix is defined as:
\be
F_{ab} = \sum_{\rm \mu \mu'} \frac{\partial S_{\rm \mu}^{\rm
T}}{\partial a}C_{\rm \mu \mu'}^{-1}\frac{\partial S^{\rm T}_{\rm \mu'}}{\partial  b}
\ee   
where $S$ is the signal described in Eq.~\ref{eq:gen_bis} and  $\rm \mu$, $\rm \mu'$, $a$ and $b$ run over 
the triplets $\{j_{1},j_{2},j_{3}\}$ and the parameter set $\{\fnl,A_{\rm D},A_{\rm F},A_{\rm S}\}$. In detail we have:
\be
F = \tiny{ \left(
\begin{array}{cccc}
S^{NG}C^{-1}S^{NG}         & S^{NG}C^{-1}S^{\rm{D}}         & S^{NG}C^{-1}S^{\rm{F}}& S^{NG}C^{-1}S^{\rm{S}} \\
S^{\rm D}C^{-1}S^{NG} & S^{\rm D}C^{-1}S^{\rm{D}} & S^{\rm D}C^{-1}S^{\rm{F}} &S^{\rm D}C^{-1}S^{\rm{S}} \\
S^{\rm {F}}C^{-1}S^{NG} & S^{\rm F}C^{-1}S^{\rm{D}} & S^{\rm F}C^{-1}S^{\rm{F}} &S^{\rm F}C^{-1}S^{\rm{S}} \\
S^{\rm {S}}C^{-1}S^{NG} & S^{\rm S}C^{-1}S^{\rm{D}} & S^{\rm S}C^{-1}S^{\rm{F}} &S^{\rm S}C^{-1}S^{\rm{S}} \\
\end{array}
\right)}\nonumber
\ee
where the sum over the triplets is implicit. 

The marginalized error on $\fnl$ arising from this Fisher analysis is  
$\Delta \fnl= 45$, very close to the limits obtained with the Monte Carlo approach;
 this confirms the efficiency of our estimation method. In
Fig.~\ref{fig:scatter} we show the 1$\sigma$ and 2$\sigma$ Fisher
error ellipses together with the Markov chain output; here each plot
presents the significance region in the case where the other
parameters are fixed at their fiducial value. It can be seen that the
free-free component seems slightly anti-correlated with  $\fnl$, the
cross-correlation coefficient being $\rho \equiv
\mbf{-}S^{NG}C^{-1}S^{\rm{F}} / ({S^{NG}C^{-1}S^{NG}\times S^{\rm
F}C^{-1}S^{\rm{F}} })^{1/2}=-0.10$. The other foreground components, i.e.\ dust and synchrotron, 
show a positive correlation, demonstrated by the orientation of the ellipses, with $\rho=0.17$ and $\rho=0.12$,
respectively.  
We can also calculate the cross-correlation between the various foreground amplitudes based on their non-Gaussianity: 
the dust-synchrotron correlation is $\rho=-0.87$, the dust-free-free correlation is $\rho=-0.29$ and the 
synchrotron-free-free correlation is $\rho=-0.58$. The scatter of the over-plotted distribution of Gaussian
simulations on the same $A_{\rm I}$-$\fnl$ plane is in 
excellent agreement with the
ellipses, which confirms again the consistency of our procedure.
We can also use the scatter in the foreground amplitudes to estimate limits on the foreground contamination using 
the bispectra; these are generally the same order as the limits arising from the fitting of the power spectrum discussed above.

As a check of our assumption that we are not strongly affected by potential 
cross-correlations among the foregrounds templates, we also considered 
the three foreground signals combined in a single foreground template. 
This effectively maps to a perfect covariance between the various foreground amplitudes.
The analysis for this combination template is identical to the individual foreground analyses 
discussed above.
We use $A_{\rm g}$ to describe the amplitude of the 
the needlet bispectrum of the total foreground map, obtained summing
the three templates. We found $\fnl = 37 \pm 43$ at 1 $\sigma$
confidence level, very close to the result when the foregrounds are included individually and assumed to be 
uncorrelated.  We also plot the two-dimension scatter plot in the
plane $\fnl$-$A_{\rm g}$ in Fig.~\ref{fig:scatter}. The correlation results $\rho\simeq0.14$,
in good agreement with the sum of the three correlations of
thermal dust, free-free and synchrotron emission, conferming the
reliability of the approach we followed. Finally, we also applied our improved estimator to the raw WMAP 5-year
temperature maps, where no foreground removal was attempted. In this case, the estimate of the parameter $\fnl$ decreases to
20: this is consistent with the overall positive value we obtain for
the correlation between foregrounds and primordial non-Gaussianity and is also consistent with earlier work, such as
\cite{Komatsu2008wmap5} and \cite{YadavWandelt2007}.

\begin{figure*}
\center
\incgr[width=\columnwidth]{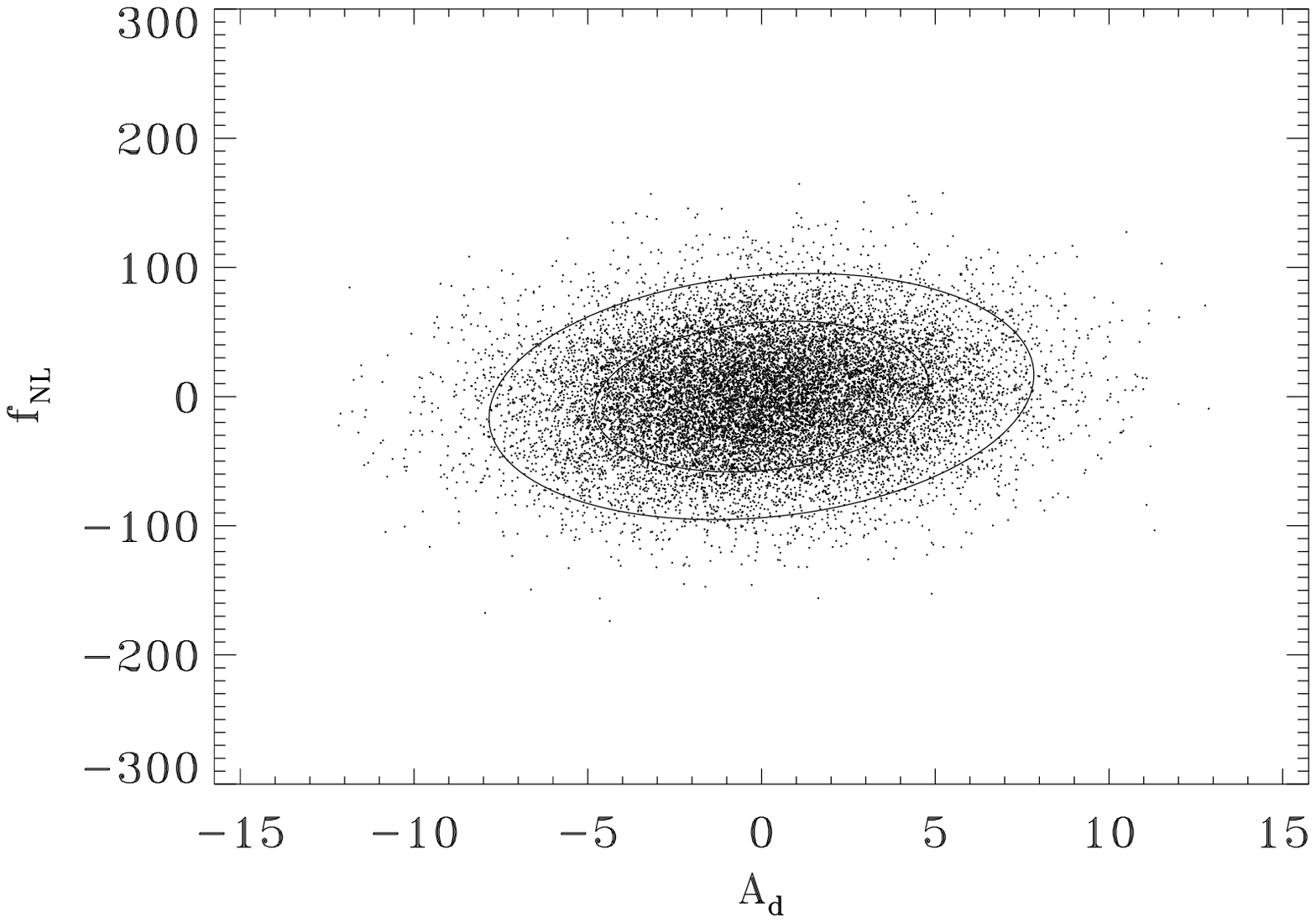}
\incgr[width=\columnwidth]{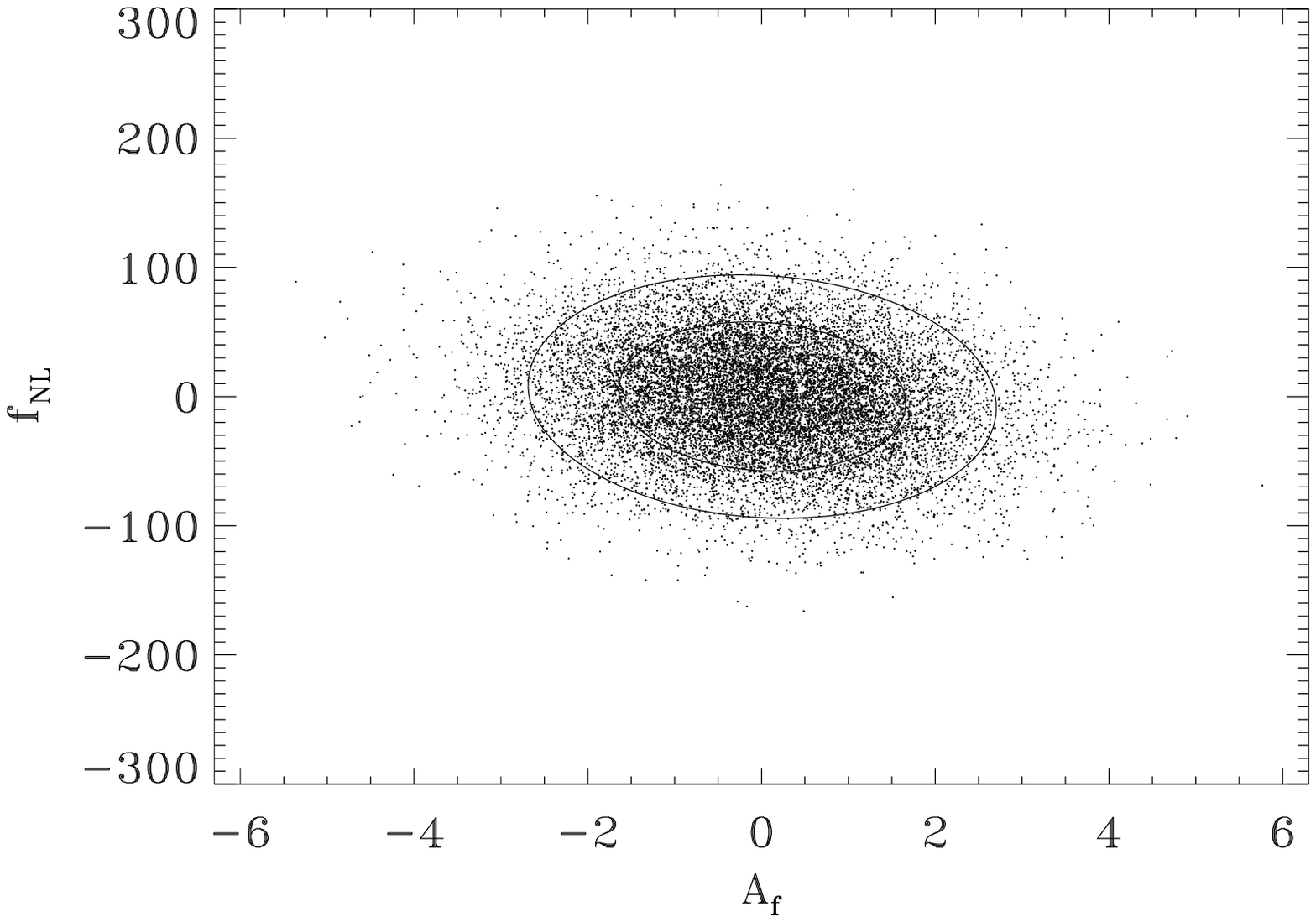}
\incgr[width=\columnwidth]{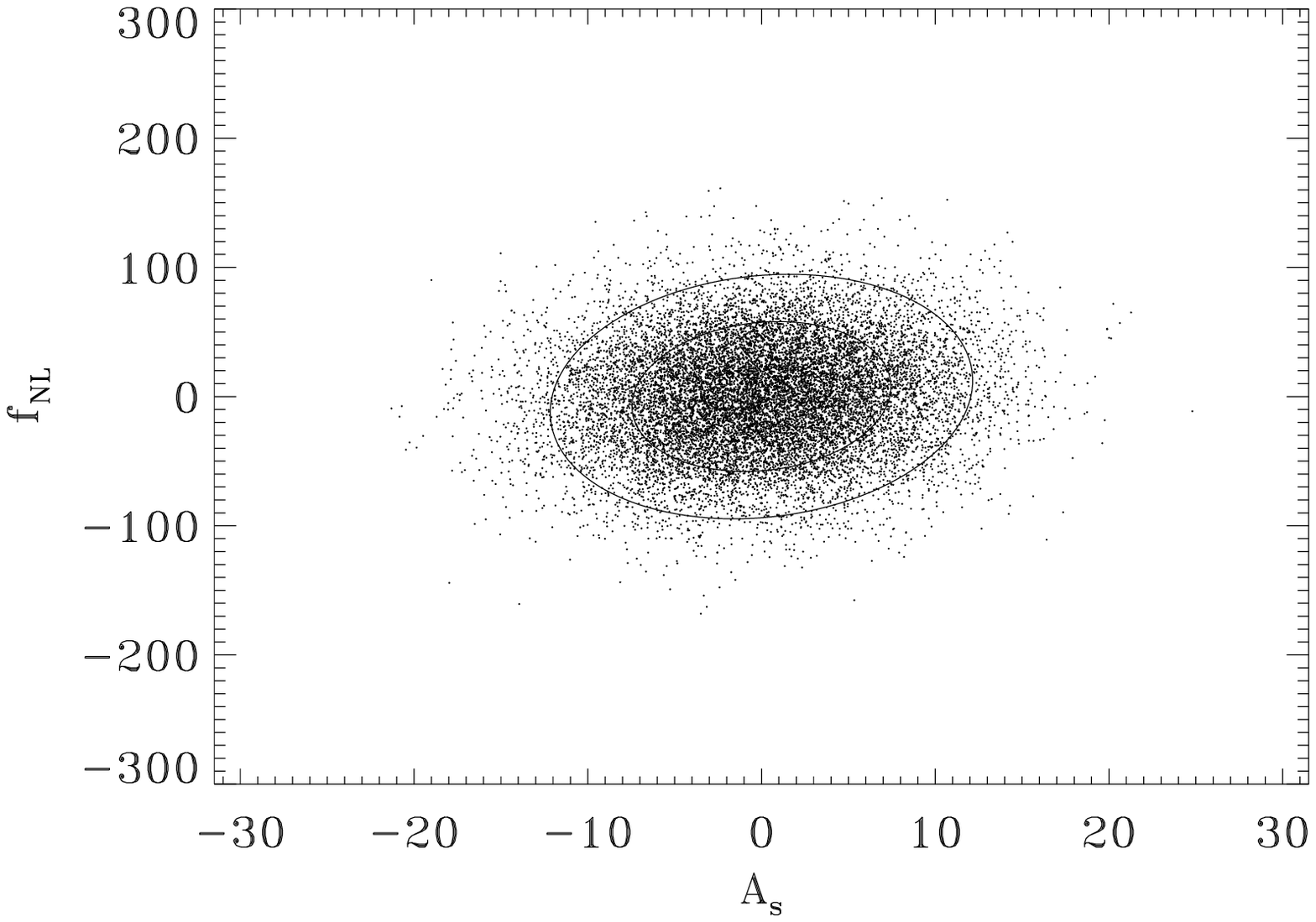}
\incgr[width=\columnwidth]{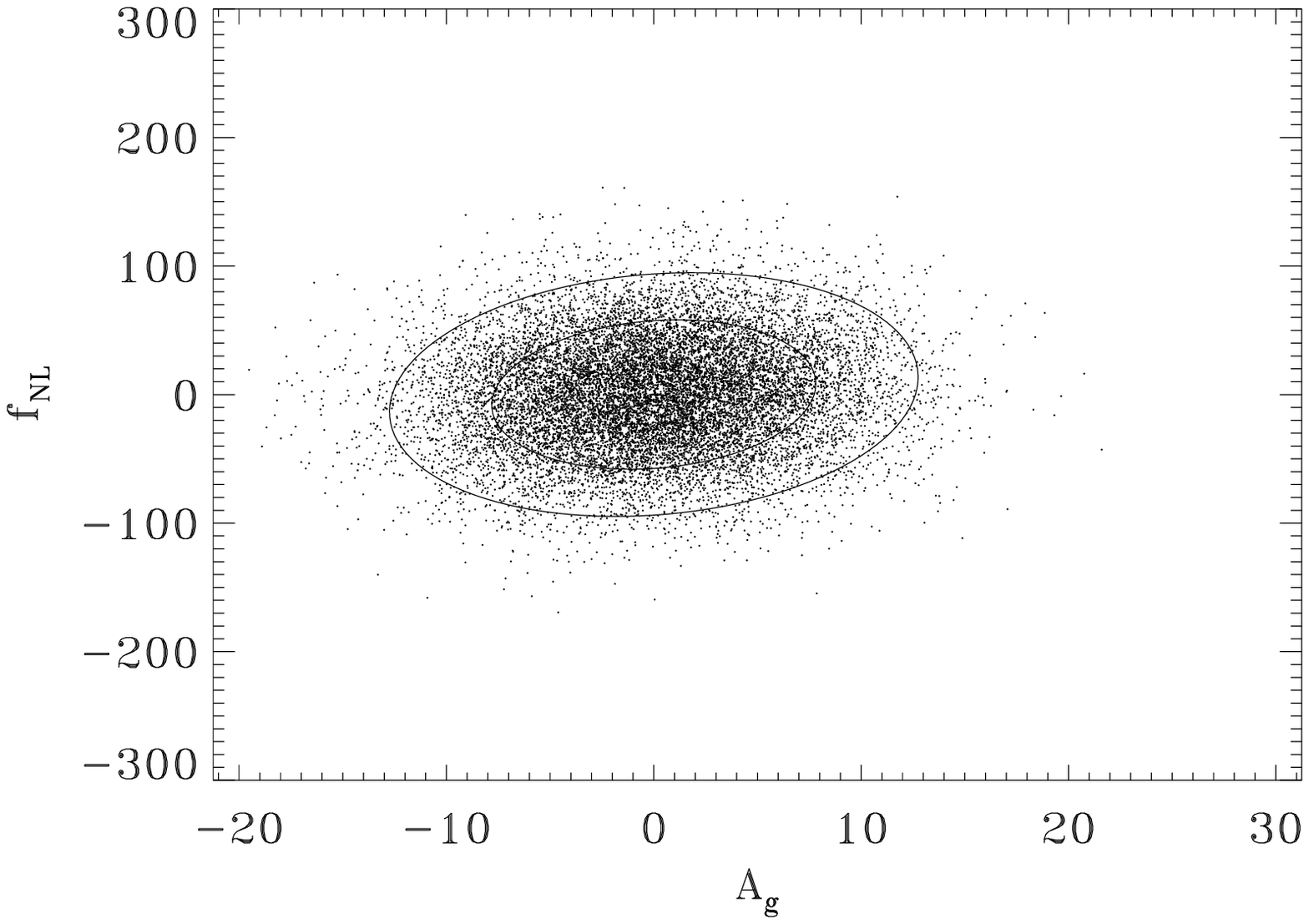}
\caption[Scatter plots for a two dimensional analysis]{Scatter plots
for two dimensional analyses, marginalising over each foreground
parameter separately.  The fourth plot show the results arising when the foreground 
is assumed to be the sum of the three templates; that is, effectively assuming the foreground amplitudes were strongly correlated with each other. 
The 1$\sigma$ and 2$\sigma$
confidence contours derived from the Fisher  analysis are
superimposed.}
\label{fig:scatter}
\end{figure*}

\section{Conclusions}
\label{sect:concl}
In this paper we have presented a procedure to marginalize the
residual foregrounds when estimating $\fnl$ in the needlet
bispectrum framework. However it is important to stress that this
algorithm does not strictly rely on needlets properties and it can be
easily applied to any linear estimator. With the foreground marginalization, 
we found, for WMAP 5-year data, that the error bars are enlarged by about $10\%$ 
with respect to the estimate obtained without marginalizing.
Foreground residuals can have different effects when different estimators are used 
to characterised  primordial non-Gaussianity.
Comparing with other foreground analyses, our results seem to go in the direction of \cite{YadavWandelt2007}, 
where they argued that foregrounds negatively biased the $\fnl$. However, \cite{Smith2009fnlFore} draw other conclusions 
showing that the sign of the biasing could depend on the choice of weighting in the method of  
estimating $\fnl$. Since needlets coefficients are essentially a rebinning of the filtered harmonic 
coefficients $a_{lm}$, this could explain the different behaviours. This issue probably needs to be examined separately for  
each test of non-Gaussianity, since different tests act differently on different spaces (e.g. harmonic, pixel, wavelet), suggesting that the influence of foregrounds on estimating $\fnl$ is not unique.       
    
All this reinforces the argument that a careful analysis with different tests of non-Gaussianity is crucial to 
discriminate between primordial non-Gaussianity and spurious effects.
Our procedure could be improved further by incorporating any covariance which may arise in the foreground subtraction methods, and by fully considering the effects arising from the potential correlations between different foreground maps.  
Such analyses will be essential for modelling the non-Gaussianity of future experiments like Planck, where the error bars on $\fnl$ are expected 
drastically reduced ($\Delta \fnl\sim 3-5$ \cite{KomatsuSpergel2001,Babich:2004}), making the uncertainties introduced by foregrounds extremely relevant.  

\subsection*{Acknowledgements}
We are grateful to Domenico Marinucci and Francesco Piacentini for fruitful discussions.
We thank Michele Liguori, Frode Hansen and Sabino Matarrese for 
providing us with the primordial non-Gaussian maps. 
The ASI contract LFI activity of Phase2 is acknowledged.
   
\bibliography{Biblio}

\end{document}